\documentclass[a4paper,aps,pre,floatfix,nofootinbib,twocolumn,showpacs]{revtex4-1}

\usepackage[pdftex]{graphicx}
\usepackage{latexsym,amsmath}
\usepackage[pdftex]{color}
\usepackage{hyperref}

\newcommand{\av}[1]{\langle #1 \rangle}

\newcommand{\eps}{\varepsilon}
\newcommand{\lT}{\lambda}

\newcommand{\FigPath}{.}

\begin{document}

\title{Topological properties of a time-integrated activity driven network}

\author{Michele Starnini} 

\affiliation{Departament de F\'\i sica i Enginyeria Nuclear,
  Universitat Polit\`ecnica de Catalunya, Campus Nord B4, 08034
  Barcelona, Spain}

\author{Romualdo Pastor-Satorras} 

\affiliation{Departament de F\'\i sica i Enginyeria Nuclear,
  Universitat Polit\`ecnica de Catalunya, Campus Nord B4, 08034
  Barcelona, Spain}

\date{\today}

\begin{abstract} 
  Here we consider the topological properties of the integrated
  networks emerging from the activity driven model [Perra \textit{at
    al.} Sci. Rep.  \textbf{2}, 469 (2012)], a temporal network model
  recently proposed to explain the power-law degree distribution
  empirically observed in many real social networks. By means of a
  mapping to a hidden variables network model, we provide analytical
  expressions for the main topological properties of the integrated
  network, depending on the integration time and the distribution of
  activity potential characterizing the model. The expressions
  obtained, exacts in some cases, the results of controlled asymptotic
  expansions in others, are confirmed by means of extensive numerical
  simulations. Our analytical approach, which highlights the
  differences of the model with respect to the empirical observations
  made in real social networks, can be easily extended to deal with
  improved, more realistic modifications of the activity driven
  network paradigm.
\end{abstract}


\maketitle

\section{Introduction}

Modern network science allows to represent and rationalize the
properties and behavior of complex systems that can be represented in
terms of a graph \cite{Newman2010,Dorogobook2010}.  Research in this
area has focused in a twofold objective: A data-driven effort to
characterize the topological properties of real networks
\cite{barabasi02,Dorogovtsev:2002,caldarelli2007sfn}, and a posterior
modeling effort, aimed at understanding the microscopic mechanisms
yielding the observed topological properties
\cite{barabasi02,Newman2010}, as well as the effects that a complex
topology has on dynamical processes running on top of it
\cite{dorogovtsev07:_critic_phenom,barratbook}.

Until recently, a large majority of work in the field of network
science has been concerned with the study of \textit{static} networks,
i.e. networks in which topological properties do not change in
time. Presently, however, a lot of attention is being devoted to the
\textit{temporal} dimension of networked systems
\cite{Holme:2011fk}. Indeed, many real networks are actually dynamical
structures, in which edges appear, vanish or are rewired at different
times scales.  An important example is given by social networks
\cite{wass94}, in which social relationships are represented by a
succession of contact or communication events, continuously created or
terminated between pairs of individuals. In this sense, the social
networks previously considered in the literature
\cite{newmancitations01,Barabasi:1999,liljeros_web_2001} represent an
projection or temporal integration of time-varying graphs, in which
all the links that have appeared at least once in a time integration
window $T$ are present in the projection.

The recent availability of large digital databases and the deployment
of new experimental infrastructures have made possible the real-time
tracking of social interactions in groups of individuals and the
reconstruction of the corresponding temporal networks
\cite{Oliveira:2005fk,Barabasi:2005uq,Gonzalez:2008fk,10.1371/journal.pone.0011596}.
The newly gathered empirical data poses new fundamental questions
regarding the properties of temporal networks, questions which have
been addressed through the formulation of theoretical models, aimed at
explaining both the temporal patterns observed and their effects on
the corresponding integrated networks
\cite{PhysRevE.83.056109,journals/corr/abs-1106-0288,citeulike:7974615,starnini_modeling_2013}.

Especially interesting in this perspective is the \textit{activity
  driven} social network model recently introduced by Perra \textit{et
  al.}  \cite{2012arXiv1203.5351P}, aimed in particular to capture the
relation between the dynamical properties of social temporal networks
and the topological properties of their corresponding time projections.
The key element in the definition of this model is the observation
that the formation of social interactions is driven by the
\textit{activity} of individuals, urging them to interact with their
peers, and by the empirical fact that different individuals show
different levels of social activity \cite{2012arXiv1203.5351P}.  Based
in the concept of \textit{activity potential}, defined as the
probability per unit time that an individual engages in a social
activity, Ref.~\cite{2012arXiv1203.5351P} proposed an activity driven
social network model, in which individuals start interactions, that
span for a fixed length of time $\Delta t$, with probability
proportional to their activity potential.  The model output is thus
given by a sequence of graphs, depending on the distribution $F(a)$ of
the activity potential, which are updated every time interval $\Delta
t$. The topological properties of the integrated activity driven
network were related in Ref.~\cite{2012arXiv1203.5351P} at the level
of the degree distribution, which, by means of approximate arguments,
was shown to be proportional to the activity potential distribution
$F(a)$. However, despite the interest of the model, expressions for
the rest of topological observables have been still lacking, a fact
that hampers its possible validation as a generator of realistic
integrated social networks, as well as the identification of the
particular role that integration time has on the behavior of dynamical
processes running on top of the temporal network
\cite{perra_random_2012,ribeiro2012quantifying}.

Here we address the study of the aggregated network generated by the
activity driven model, obtained by integrating the temporal network up
to a given time $T$, by considering a mapping of the integrated
network to a hidden variables model \cite{PhysRevE.68.036112}
depending on the activity potential distribution and the considered
time $T$.  We obtain a set of expressions for the degree distribution,
degree-degree correlations and clustering coefficient of the
aggregated network that are exact in the limit of large network size
$N \to \infty$ and finite time $T$, and which are amenable to analytic
asymptotic expansions in this same limit. The expressions obtained,
confirmed by numerical simulations, corroborate the basic assumption
of the activity driven model linking social activity with network
topology. Moreover, the formalism proposed can be extended to
generalizations of the activity driven model, opening thus the path to
the analytical solution of model extensions, aimed at better
reproducing the topological features of real integrated social
networks.

We have organized the paper: Section~\ref{sec:hidd-vari-form} defines
the activity driven network model. Sec.~\ref{sec:hidd-vari-form} gives
a brief review of the hidden variables network formalism;
Sec.~\ref{sec:mapp-integr-netw} deduces the mapping of the activity
driven model onto this formalism.  In Sec.~\ref{sec:topol-prop-binary}
we compute the topological properties of the integrated activity
network model as a function of time $T$, providing exact expressions
as well as asymptotic results in the limit $T/N$ small, for different
forms of the activity potential distribution.  In
Sec.~\ref{sec:model-extensions} we discuss the extension of the
formalism presented to possible variations of the activity driven
model. Finally, we summarize our results and conclusions in
Sec.~\ref{sec:conclusions}.

\section{The activity driven network model}
\label{sec:activity-model}

The activity driven network model proposed in
Ref.~\cite{2012arXiv1203.5351P} is defined in terms of $N$ individuals
$i$ (agents), each one of them characterized by her activity potential
$a_i$, defined as the probability that she engages in a social
act/connection with other agents per unit time. The activity of the
agents is a (quenched) random variable, extracted from the activity
potential distribution $F(a)$, which can take a priori any form.  The
model is defined by means of a synchronous update scheme, time being
measured in units of the life span of each connection $\Delta t$. It
proceeds by creating a succession of instantaneous networks
$\mathcal{G}_t$, $t=0, \Delta t, 2 \Delta t, \ldots, n \Delta t,
\ldots$ At a given time $t$, all previous edges are deleted and we
start with $N$ disconnected individuals. Each one of them is checked
and becomes active with probability $a_i \Delta t$. Active agents
generate $m$ links (start $m$ social interactions) that are connected
to $m$ other agents selected uniformly at random. Finally, time is
updated as $t \to t + \Delta t$. This procedure implies that all edges
in the temporal network have the same constant time duration $\Delta
t$.  In order to avoid complications due to the differences in the
number of emitted and received connections arising from using a
synchronous approach\footnote{Indeed, in a synchronous scheme, every
  time step an agent fires at most one connection, but can receive a
  number $n$ of connections, given trivially by a binomial
  distribution.}, here we consider a probabilistic recipe for the
instantaneous network construction: Each microscopic time step $\Delta
t$, we choose $N$ agents, uniformly at random, and check sequentially
each one of them for activation and eventual link emission.  We avoid
self and multiple connections.

To simplify the analytical calculations performed below, in the
following we choose $\Delta t= m =1$. Both quantities can be however
restored by a simple rescaling of the activity potential and the
integration time $T$. We notice that imposing $\Delta t=1$ implies
restricting the activity potential to be probability, and thus to be
limited in the interval $a \in [0,1]$.

\section{Hidden variables formalism: A short review}
\label{sec:hidd-vari-form}

The class of network models with hidden variables was introduced in
Ref.~\cite{PhysRevE.68.036112} (see also
\cite{PhysRevLett.89.258702,Soderberg:2002fk}) as a generalization of
the random network Gilbert model \cite{Dorogobook2010}, in which the
probability of connecting two vertices is not constant, but depends on
some intrinsic properties of the respective vertices, their so-called
hidden variables. This class of models is defined as follows: Starting
from a set of $N$ disconnected vertices and a general hidden variable
$h$, we construct an undirected network with no self-edges nor
multiple connections, by applying these two rules:
\begin{enumerate}
\item To each vertex $i$, a variable $h_i$ is assigned, drawn at
  random from a probability distribution $\rho(h)$.
\item For each pair of vertices $i$ and $j$, $i \neq j$, with hidden
  variables $h_i$ and $h_j$, respectively, an edge is created with
  probability $r(h_i, h_j)$, the connection probability, which is a
  symmetric function bounded by $0 \leq r(h, h') \leq 1$.
\end{enumerate}
Each model in the class is fully defined by the functions $\rho(h)$
and $r(h, h')$, and all its topological properties can be derived as a
function of these two parameters. These topological properties are
encoded in the propagator $g(k|h)$, defined as the conditional
probability that a vertex with hidden variable $h$ ends up connected
to $k$ other vertices. The propagator is a normalized function,
$\sum_k g(k|h) =1$, whose generating function $\hat{g}(z|h) = \sum_k
z^k g(k|h)$ fulfills the general equation \cite{PhysRevE.68.036112}
\begin{equation}
  \ln   \hat{g}(z|h) = N \sum_{h'} \rho(h') \ln \left[ 1-(1-z) r(h, h')
  \right].
  \label{eq:3} 
\end{equation}
From this propagator, expressions for the topological
properties of the model can be readily obtained
\cite{PhysRevE.68.036112}:
\begin{itemize}
\item Degree distribution:
  \begin{equation}
    P(k)=\sum_h g(k|h)\rho(h).
    \label{eq:pk}
  \end{equation}

\item Degree correlations, as measured by the average degree of the
  neighbors of the vertices of degree $k$, $\bar{k}^{nn}(k)$
  \cite{alexei}:
  \begin{equation}
    \bar{k}^{nn}(k) = 1+ \frac{1}{P(k)} \sum_h \rho(h) g(k|h)
    \bar{k}^{nn}(h),
    \label{eq:13}
  \end{equation}
  where we have defined
  \begin{equation}
    \bar{k}^{nn}(h) = \frac{N}{\bar{k}(h)}  \sum_{h'}  \rho(h') \bar{k}(h')
    r(h,h'),
    \label{eq:12}
  \end{equation}
  and
  \begin{equation}
    \bar{k}(h) = N \sum_{h'} \rho(h') r(h,h'),
    \label{eq:5}
  \end{equation}
  which is the average degree of the vertices with hidden variable
  $h$.

\item Average clustering coefficient $\av{c}$, defined
  as the probability that two vertices are connected, provided that
  they share a common neighbor \cite{watts98}
  \begin{equation}
    \av{c} = \sum_h \rho(h) \bar{c}(h),
    \label{eq:6}
  \end{equation}
 where we have defined
  \begin{equation}
    \label{eq:24}
    \bar{c}(h) = \sum_{h', h''} p(h'|h) r(h', h'') p(h''|h),
  \end{equation}
  and
  \begin{equation}
    \label{eq:32}
    p(h'|h) = \frac{N \rho(h') r(h, h')}{ \bar{k}(h)}.
  \end{equation}
  Additionally, one can define the clustering spectrum, as measured by
  the average clustering coefficient of the vertices of degree $k$,
  $\bar{c}(k)$ \cite{alexei,ravasz_hierarchical_2003}
  \begin{equation}
    \label{eq:23}
    \bar{c}(k) = \frac{1}{P(k)}  \sum_h \rho(h) g(k|h) \bar{c}(h),
  \end{equation}

\end{itemize}

\section{Mapping the integrated network to a hidden variables model}
\label{sec:mapp-integr-netw}

The activity driven network model generates a time series of
instantaneous sparse networks, with an average degree $\av{k}_t \simeq
2 \av{a}$, where $\av{a} = \sum_a a F(a)$. The integrated network at
time $T$ is constructed by performing the union of the instantaneous
networks, i.e. $\mathcal{G}_T = \cup_{t=0}^T \mathcal{G}_t$. In this
integrated network, vertices $i$ and $j$ will be joined by an edge if
there has ever been a connection created between them in any of the
instantaneous networks at $0 \leq t \leq T$. The key point to map the
integrated network to a hidden variables model resides in computing
the probability $\Pi_T(i,j)$ that two vertices $i$ and $j$ become
eventually joined at time $T$. This probability is given by
$\Pi_T(i,j) =1- Q_T(i,j)$, where $Q_T(i,j)$, the probability that no
connection has ever been created between agents $i$ and $j$ up to time
$T$, can be calculated as follows: At time $T$, an agent $i$ will have
become active $z$ times with probability $P_T(z)$.  Given the
definition of the model, at time $T$ we have selected $T N$ agents to
check for activation. The number of times $z$ that agent $i$ has
become active will be given by the binomial distribution
\begin{equation}
  \label{eq:21}
  P_T(z) = \binom{T N }{z} \left(
    \frac{a_i}{N}\right)^{z} 
  \left(1- \frac{a_i}{N}\right)^{TN  -z},
\end{equation}
and analogously for agent $j$.  Now, vertices $i$ and $j$ will be
connected in the integrated network if at least one of the links
generated from $i$ reaches $j$, or vice-versa. Since every time that
she becomes active, an agent creates a connection targeted to a
randomly chosen peer, the probability $Q_T(i,j)$ is given by
\begin{eqnarray}
  Q_T(i,j) &=& \sum_{z_i,z_j}  P_T(z_i)  P_T(z_i)
  \left(1-\frac{1}{N}\right)^{ z_i} 
  \left(1-\frac{1}{N}\right)^{ z_j} \nonumber\\
  &=& 
  \left[ \left( 1 - \frac{a_i}{N^2}
      \right)
    \left( 1 - \frac{a_j}{N^2}
       \right)
  \right]^{T N }, 
  \label{eq:10}
\end{eqnarray}
where we have performed the summation using the probability
distribution in Eq.~\eqref{eq:21}. We see now that the probability
that agents $i$ and $j$ are connected in the integrated network at
time $T$ depends only on their respective activity potentials $a_i$
and $a_j$, which are random variables with distribution $F(a)$.  The
mapping to a hidden variables network is thus transparent:
\begin{itemize}
\item Hidden variable: $h \to a$.
\item Distribution of hidden variables: $\rho(h) \to F(a)$.
\item Connection probability: $r(h, h') \to \Pi_T(a, a')$.
\end{itemize}

At very large times, the integrated network emerging from the activity
driven model will trivially tend to a fully connected
network. Interesting topology will thus be restricted to the limit of
small $T$ compared with the network size $N$. In this limit,
Eq.~(\ref{eq:10}) can be simplified, yielding
\begin{eqnarray}
  \Pi_T(a,a') & = & 1-Q_T(a,a') \simeq1 - \left[ 1 - \frac{(a + a')}{N^2}
  \right]^{TN} \nonumber \\ 
  &\simeq & 1- \exp\left[ -
    \lT (a +a') \right],
  \label{eq:7}
\end{eqnarray}
where we have neglected terms of order $\mathcal{O}(N^{-2})$ and
defined the parameter
\begin{equation}
  \label{eq:9}
  \lambda =  \frac{T}{N}.
\end{equation}
An explicit calculation of the connection probability for a factor
$m>1$ and a time interval $\Delta t \neq 1$ can be easily performed;
in the limit of large $N$ and constant $\lambda$, the only change
ensuing is a rescaling of time, $T \to Tm$, the value of $\Delta t$
becoming canceled in the process of taking the limit $ \lambda\to0$.

\section{Topological properties of the integrated activity driven
  network}
\label{sec:topol-prop-binary}

Here we will apply the formalism presented in
Sec.~\ref{sec:hidd-vari-form} to provide analytic expressions
characterizing the topology of the integrated network resulting from
the activity driven model.  For the sake of concreteness, we will
focus in the following activity potential distributions, in the
continuous $a$ limit:
\begin{itemize}
\item Constant activity:
  \begin{displaymath}
    F(a) = \delta_{a, a_0} , \;\; \mathrm{with} \; 0< a_0 < 1.
  \end{displaymath}
\item Homogeneous activity:
  \begin{equation}
    F(a) = 1/a_{\max}, \;\; \mathrm{with} \;  0 \leq a \leq a_{\max}
    \leq 1. 
    \label{eq:2}
  \end{equation}
\item Power-law distributed activity:  
  \begin{equation}
    \label{eq:11}
     F(a) = (\gamma-1) \eps^{\gamma-1} a^{-\gamma}, \;\; \mathrm{with} \; a \in
    [\eps,1].
  \end{equation}
\end{itemize}
In the last case, where we consider $\gamma>2$, in accordance with
experimental evidence \cite{2012arXiv1203.5351P}, we have introduced a
lower cut-off $0<\eps \ll 1$ in order to avoid dangerous divergences
in the vicinity of zero.

\subsection{Degree distribution}
\label{sec:degree-distribution}

In order to compute the degree distribution, we have to solve and
invert the generating function equation Eq.~\eqref{eq:3}, an almost
impossible task to perform exactly, except in the case of very simple
forms of the activity potential distribution. So, in the case of
constant activity, $F(a) = \delta_{a, a_0}$, we have
\begin{equation}
  \label{eq:58}
   \hat{g}(z|a_0) = \left[ z \Pi_T(a_0,a_0) + (1-\Pi_T(a_0,a_0)) \right]^N,
\end{equation}
which corresponds to the generating function of a binomial
distribution \cite{Wilf:2006:GEN:1204575}. Therefore, in the limit of
large $N$ and constant $\lT$, the degree distribution takes the
Poisson form
\begin{equation}
  \label{eq:59}
  P_T(k) = e^{-\mu} \frac{\mu^k}{k!}
\end{equation}
with parameter $\mu = N \left(1- e^{-2 \lT a_0}\right)$, which, for
fixed $T$ and large $N$, can be approximated as $\mu \simeq 2 T
a_0$. 

For a nontrivial activity distribution $F(a)$, we must resort to
approximations. We therefore focus in the interesting limit of small
$\lT$, which corresponds to fixed $T$ and large $N$, which is the one
yielding a non-trivial topology.  In this limit, we can approximate
the connection probability as
\begin{equation}
  \label{eq:1}
  \Pi_T(a,a') \simeq \lT(a +a').
\end{equation}
Introducing this expression into Eq.~\eqref{eq:3} and performing a new
expansion at first order in $\lambda$, we obtain
\cite{PhysRevE.68.036112}
\begin{eqnarray}
   \ln   \hat{g}(z|a) 
  &\simeq&  (1-z) \lT N \sum_{a'} F(a') (a+a') \\
  &=&  (1-z) \lT N (a+\av{a}).  \label{eq:28}
\end{eqnarray}
The generating function of the propagator is a pure exponential, which
indicates that the propagator itself is a Poisson distribution
\cite{Wilf:2006:GEN:1204575}, i.e.
\begin{equation}
  \label{eq:30}
  g(k|a) = e^{-T (a+\av{a})} \frac{\left[ T (a+\av{a})
    \right]^k}{\Gamma(k+1)},
\end{equation}
where $\Gamma(x)$ is the Gamma (factorial) function \cite{abramovitz}.
From Eq.~\eqref{eq:pk} we obtain the general expression for the degree
distribution
\begin{equation}
  \label{eq:pk_general}
  P_T(k) =  \frac{T^k}{\Gamma(k+1)} \sum_a
    F(a) \left[a+\av{a} \right]^k e^{-T (a+\av{a})}.
\end{equation}

In the case of a homogeneous activity distribution, $F(a)=
{a_{\max}}^{-1}$, for which $\av{a} = a_{\max}/ 2$, we can integrate
directly Eq.~(\ref{eq:pk_general}), to obtain
\begin{equation}
  \label{eq:57}
   P_T(k) =  \frac{\Gamma\left(k+1, T \av{a} \right) -
     \Gamma\left(k+1, 3T \av{a} \right)}{2T\av{a} \; \Gamma(k+1)}.
\end{equation}
where $\Gamma(x,z)$ is the incomplete Gamma function
\cite{abramovitz}.  

More complex forms of the activity distribution do not easily yield to
an exact integration, and more approximations must be performed. In
particular, the asymptotic form of the degree distribution can be
obtained by performing a steepest descent approximation. Thus, we can
write
\begin{equation}
  \label{eq:31}
  P_T(k) =  \frac{1}{\Gamma(k+1)}  \int F(a) e^{\phi(a)} da,
\end{equation}
where we have defined
\begin{equation}
  \label{eq:16}
  \phi(a) = k \ln[T(a + \av{a})] - T(a+\av{a}).
\end{equation}
The function $\phi(a)$ has a sharp maximum around $a_M = \frac{k}{T} -
\av{a}$. Performing a Taylor expansion up to second order, we can
write $\phi(a) \simeq \phi(a_M) - \frac{T^2}{2 k}
\left[a-a_M\right]^2$, with $ \phi(a_M) = k \ln(k) - k$. Now, for $T^2
/ k \gg 1$, the function $e^{- \frac{T^2}{2 k} \left[a-a_M\right]^2}$
is strongly peaked around the maximum $a_M$; therefore we can
substitute the activity potential by its value at the maximum, to
obtain
\begin{eqnarray}
  P_T(k) &\simeq&  \frac{e^{\phi(a_M)}F(a_M)
   }{\Gamma(k+1)}  \int_{-\infty}^{\infty} e^{- \frac{T^2}{2 k}
     \left[a- a_M\right]^2} da \nonumber\\
   &=&  \frac{ \sqrt{2 \pi k} k^k e^{-k}
   }{T \Gamma(k+1)} F\left(\frac{k}{T} - \av{a}\right),
  \label{eq:17}
\end{eqnarray}
where we have extended the integration limits to plus and minus
infinity. In the large $k$ limit, we can use Stirling's approximation,
$\Gamma(k+1) \sim \sqrt{2 \pi k} k^k e^{-k}$, to obtain the asymptotic
form
\begin{equation}
  \label{eq:18}
  P_T(k) \sim \frac{1}{T} F\left(\frac{k}{T} -
    \av{a}\right). 
\end{equation}
In this expression we recover, using more rigorous arguments, the
asymptotic form of the integrated degree distribution obtained in
Ref.~\cite{2012arXiv1203.5351P}.
The limits of validity of this expression are however now transparent, 
being explicitly $N \gg T \gg 1$ and $T^2 \gg k \gg 1$.

\begin{figure}[t]
  \includegraphics[height=6 cm]{\FigPath/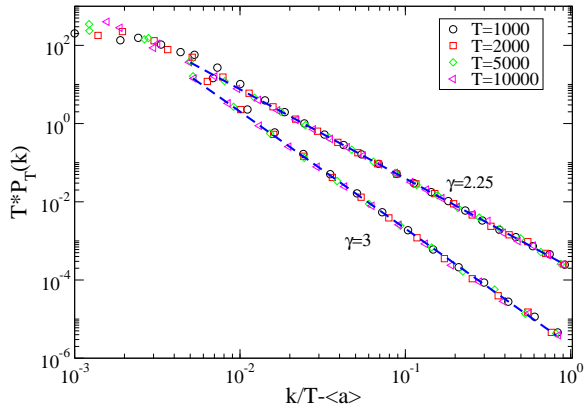}
  \caption{(color online) Rescaled degree distribution $P_T(k)$ for
    integrated networks corresponding to different values of $T$, with
    power law activity distribution with exponents $\gamma = 3$ and
    $2.25$. Network size $N=10^6$.  The behavior predicted by
    Eq.~(\ref{eq:18}) is represented as dashed lines.}
  \label{fig:k_distr1}
\end{figure}

\begin{figure}[t]
  \includegraphics[height=6 cm]{\FigPath/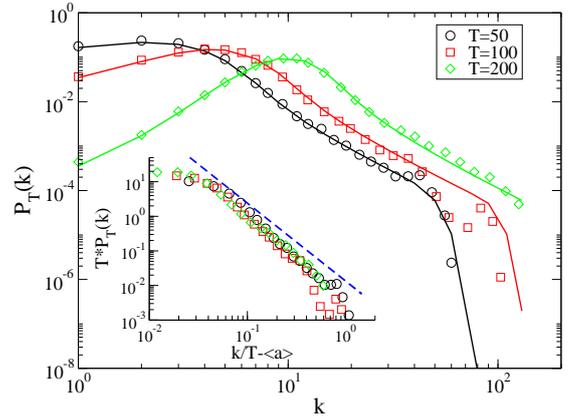}
  \caption{(color online) Degree distribution $P_T(k)$ for integrated
    networks corresponding to different values of $T$, with power law
    activity distribution with exponent $\gamma = 2.25$. Network size $N=10^3$.
    The result of a numerical integration of Eq.~(\ref{eq:pk_general})
    is showed as continuous lines.  Inset: Rescaled $P_T(k)$ shown
    against Eq.~(\ref{eq:18}), dashed in blue. }
  \label{fig:k_distr2}
\end{figure}

For the case of constant activity, $F(a) = \delta_{a, a_0}$, the
asymptotic form of the degree distribution is ${P_T(k) \sim \delta_{k,
    T a_0} / T}$, while the exact form is a Poisson distribution
centered at $2 T a_0$.  For a uniform activity, on the other hand, the
asymptotic prediction is a flat distribution, while the exact
expression can be quite different, in particular for large and small
values of $k$, see Eq.~(\ref{eq:57}).  For the case of a power-law
distributed activity, in Fig.~\ref{fig:k_distr1} we plot the degree
distribution $P_T(k)$ of the aggregated network at different values of
$T$ for networks of size $N=10^6$ and two different values of
$\gamma$.  As we can see, for such large networks sizes and values of
$\lambda \sim 10^{-2} - 10^{-3}$, the asymptotic expression
Eq.~\eqref{eq:18} represents a very good approximation to the model
behavior.  In Fig.~\ref{fig:k_distr2} we plot the degree distribution
for a smaller network size $N=10^3$. As one can see, a numerical
integration of Eq.~(\ref{eq:pk_general}) recovers exactly the behavior
of $P_T(k)$ even for small values of $k$.  With such small network
size, however, the asymptotic prediction of Eq.~\eqref{eq:18} is less
good, as shown in the inset of Fig.~\ref{fig:k_distr2}.

\subsection{Degree correlations}
\label{sec:degree-correlations}

We start from Eq.~\eqref{eq:5}, which takes the form, as a function
of time
\begin{equation}
\bar{k}_T(a) = N[1-e^{-\lT a} \Psi(\lT)],
\end{equation}
where $\Psi(\lambda)$ is the Laplace transform 
\begin{equation}
   \Psi(\lambda) \equiv  \sum_{a} F(a) e^{- \lambda a}.
\end{equation}
We can now use Eq.~\eqref{eq:12}, which leads to the exact expression
 \begin{equation}
  \label{eq:40}
   \bar{k}^{nn}_T(a) =
  N \left\{ 1 -  \Psi(\lT)\frac{\Psi(\lT)-\Psi(2 \lT) e^{-\lT
        a}}{1-\Psi(\lT) e^{-\lT a}}  
   \right\}.
\end{equation}
In order to obtain an explicit expression for $\bar{k}^{nn}_T(k)$ we
must perform the integral in Eq.~\eqref{eq:13}. In the case of a
constant activity potential, $F(a) = \delta_{a,a_0}$, we have $P_T(k)
= g(k|a_0)$. Since in this case $\Psi(\lT) = e^{-\lT a_0}$, we have
\begin{equation}
  \label{eq:14}
  \bar{k}^{nn}_T(k) = 1 + N \left[ 1 - e^{-2\lT a_0}  \right] \simeq 1 +
  2 T a_0,
\end{equation}
where the last expression corresponds to the limit of small $\lT$.
This function is independent of $k$, indicating that the integrated
network corresponding to constant activity potential has no degree
correlations. 

For more complex forms of $F(a)$, we resort to an expansion in powers
of $\lT$ to obtain an approximate expression, which at lowest order
takes the form
\begin{equation}
  \label{eq:42}
  \bar{k}^{nn}_T(a) \simeq \frac{\lT N }{a+\av{a}} \left[ \av{a^2} +
  \av{a}^2 + 2a \av{a} \right].
\end{equation}
Inserting this expression into Eq.~\eqref{eq:13}, and considering the
Poisson form of the propagator Eq.~\eqref{eq:30}, we can write
\begin{eqnarray}
  \label{eq:4}
  \bar{k}^{nn}_T(k) &\simeq & 1 + \frac{T^2 (\av{a^2} + \av{a}^2)}{k P(k)}  \int da F(a) g(k-1|a) + \nonumber \\ \nonumber
                          &           &  \frac{2T^2 \av{a}}{k P(k)}  \int da a F(a) g(k-1|a) \\  \nonumber
  &\simeq& 1 +  T^2 \frac{P(k-1)}{k P(k)} \left[ \sigma_a^2 + 2\av{a} \left(\frac{k}{T} \right) \right],  
\end{eqnarray}
where in the last expression we have performed the steepest descent
approximation used to obtain Eq.~\eqref{eq:17}, and $\sigma_a^2 =
\av{a^2} - \av{a}^2$ is the variance of the activity potential
$F(a)$. In the limit of large $k$, where $P(k-1) / P(k) \sim 1$, we
have the general form for the degree correlations
\begin{equation}
  \frac{ \bar{k}^{nn}_T(k) -1} {T}    \simeq  2\av{a}  + \sigma_a^2
  \left( \frac{ k }{ T} \right)^{-1 } .
  \label{eq:15}
\end{equation}
This expression recovers in a natural way the exact result for
constant activity potential, where $\sigma_a^2 = 0$.  From
Eq.~(\ref{eq:15}) we conclude that, in general, for an non-constant
activity distribution, the integrated networks resulting from the
activity driven model show disassortative mixing by degree
\cite{PhysRevLett.89.208701}, with a $\bar{k}^{nn}_T(k)$ function
decreasing as a function of $k$. This disassortative behavior, which
can be however quite mild in the case of small variance $\sigma_a$, as
in the case of a power law distributed activity with small $\eps$, is
in any case at odds with the assortative form observed for degree
correlations in real social networks \cite{Newman2010}.

\begin{figure}[tbp]
  \includegraphics[height=6 cm]{\FigPath/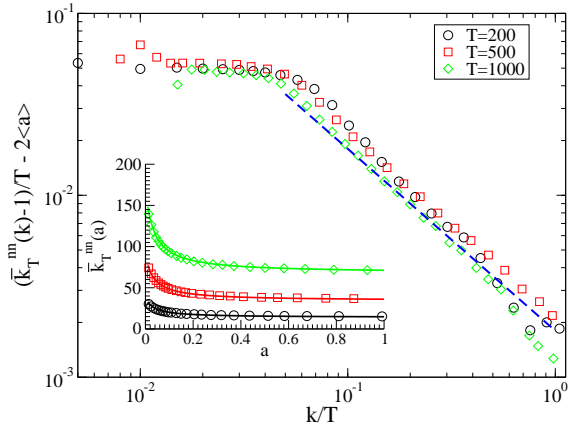}
  \caption{(color online) Main: Rescaled average degree of the nearest
    neighbors of the vertices of degree $k$, $\bar{k}^{nn}_T(k)$, for
    the integrated network with size $N=10^4$ and power law activity
    distribution with $\gamma = 2.5$, for different values of $T$.
    The prediction of Eq.~(\ref{eq:15}) is dashed in blue.  Inset:
    Average degree of the neighbors of the vertices with activity $a$,
    $\bar{k}^{nn}_T(a)$, for the same integrated network.  The
    predictions from Eq.~(\ref{eq:42}) are shown as continuous lines.
  }
  \label{fig:k_nn_k}
\end{figure}

In Fig.~\ref{fig:k_nn_k} we check the validity of Eq.~(\ref{eq:42})
and the asymptotic form Eq.~(\ref{eq:15}) in the case of power law
distributed activity. We observe that the prediction of
Eq.~(\ref{eq:42}) recovers exactly the model behavior, also in the
case of small activity $a$ (shown in the inset).  The degree
correlation, $\bar{k}^{nn}_T(k)$, is also correctly captured by the
asymptotic form Eq.~(\ref{eq:15}). Note however that, since the
variance $\sigma_a$ is small (of order $\eps^{\gamma-1}$ for
$\gamma<3$ and order $\eps^2$ for $\gamma>3$), the net change in the
average degree of the neighbors is quite small, and the integrated
network can be considered as approximately uncorrelated without
incurring in a gross error.

\subsection{Clustering coefficient}
\label{sec:clust-coeff}

The expression of the clustering spectrum at time $T$, $\bar{c}_T(k)$,
takes the form, from Eq.~\eqref{eq:23}
\begin{equation}
  \label{eq:29}
  \bar{c}_T(k) = \frac{1}{P_T(k)} \sum_a F(a) g(k|a) \bar{c}_T(a). 
\end{equation}
Using Eqs.~\eqref{eq:24}, \eqref{eq:32} and the expression for
$\bar{k}(a)$, we can write the exact form
\begin{equation}
  \label{eq:33}
  \bar{c}_T(a) = 1 - \left[\frac{\Psi(\lT) - e^{-\lT a} \Psi(2\lT)}{1 -
      e^{-\lT a} \Psi(\lT)} \right]^2.
\end{equation}
Again in the simplest case of a constant activity potential, $F(a) =
\delta_{a, a_0}$, we have $\bar{c}_T(a) = 1 - e^{-2 \lT a_0}$, which
leads to a clustering spectrum
\begin{equation}
  \label{eq:34}
  \bar{c}_T(k) \equiv \av{c}_T = 1 -  e^{-2 \lT a_0} \simeq \frac{2 T
    a_0}{N},
\end{equation}
where the last expression is valid for small $\lT$.  The clustering
spectrum is in this case constant, and equal to the average clustering
coefficient. For fixed time $T$, it is inversely proportional to the
network size, in correspondence to a purely random network. It
increases with $T$, saturating at $\av{c}_\infty =1$ for a fully
connected network in the infinite time limit.

\begin{figure}[tb]
  \includegraphics[height=6 cm]{\FigPath/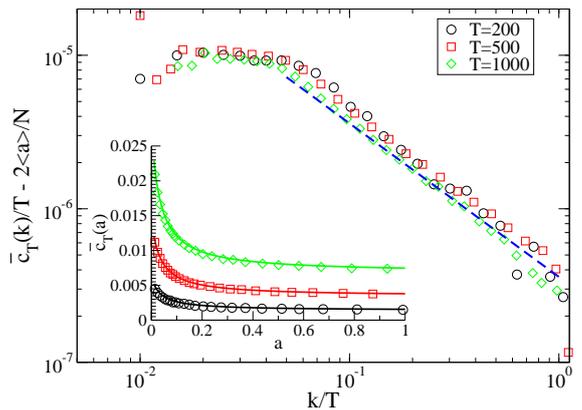}
  \caption{(color online) Main: Rescaled clustering coefficient of the
    nodes of degree $k$, $c_T(k)$, of the integrated network with size
    $N=10^4$ and power law activity distribution with $\gamma = 2.5$,
    for different values of $T$.  The prediction of Eq.~(\ref{eq:37})
    is dashed in blue.  Inset: Clustering coefficient of the nodes
    with activity $a$, $c_T(a),$ of the same integrated network.  The
    predictions from Eq.~(\ref{eq:35}) are shown as continuous lines.
  }
  \label{fig:ck}
\end{figure}

For a general activity potential distribution, we need to perform
again an expansion in $\lT$, which in this case takes the form, at
first order in  $\lT$,
\begin{equation}
  \label{eq:35}
  \bar{c}(a) \simeq \frac{2 \lT}{a + \av{a}} [\av{a^2} + a \av{a}].
\end{equation}
Inserting this form into Eq.~\eqref{eq:29}, and performing the same
steepest descend approximation applied in Eq.~\eqref{eq:4}, we obtain
\begin{equation}
\bar{c}(k) \simeq  \frac{ 2 T^2}{N}  \frac{P(k-1)}{k P(k)}  
\left[ (\av{a^2} - \av{a}^2) +  \av{a} \left(\frac{k}{T} \right) \right].
\end{equation}
In the limit of large $k$, we obtain the general form of the
clustering spectrum, valid for any activity potential,
\begin{equation}
  \label{eq:37}
  \frac{ \bar{c}(k)}{T} \simeq \frac{2 \av{a}}{N} +  \frac{2 \sigma_a^2}{N} \left( \frac{ k }{ T} \right)^{-1 }.
\end{equation}
In Fig.~\ref{fig:ck} we plot the clustering coefficient as a function
of the degree (main) and the activity (inset), in the case of power
law distributed activity. We observe that both Eq.~(\ref{eq:35}) and
Eq.~(\ref{eq:37}) recover correctly the clustering coefficient
behavior.

\section{Model extensions}
\label{sec:model-extensions}

The potency of the hidden variables formalism we have introduced above
to solve the activity driven model allows to easily extended it to
tackle the analysis of generalized models inspired in the same
principles. We can consider, indeed, different rules for activation
and reception of connections. The only limitation to be imposed in
order to properly implement the formalism is that connection rules
must be local, i.e. involving only properties of the emitting and
receiving agents.  As a simple example, we consider a sort of
``inverse'' activity driven model, in which every agent $i$ becomes
active with the same constant probability $a_i = a_0$ and, when
active, she sends a connection to another agent $j$, chosen at random
with probability proportional to some (quenched) random quantity
$b_j$, the \textit{attractiveness} of the node, i.e. with probability
$b_j / \av{b}N$. In this case, one can easily repeat the steps of the
mapping presented in Sec.~\ref{sec:mapp-integr-netw}: The number of
times $z$ that agent $i$ becomes active is now
\begin{equation}
\label{eq:19}
  P'_T(z) = \binom{T N }{z} \left(
    \frac{a_0}{N}\right)^{z} 
  \left(1- \frac{a_0}{N}\right)^{TN  -z},
\end{equation}
and the probability that $i$ and $j$ never become connected up to
time $T$ is
\begin{eqnarray}
  Q'_T(i,j) &=& \sum_{z_i,z_j}  P'_T(z_i)  P'_T(z_j)
  \left(1-\frac{b_j}{\av{b}N}\right)^{ z_i} 
  \left(1-\frac{b_i}{\av{b}N}\right)^{ z_j} \nonumber \\
  &=& 
  \left[ \left( 1 - \frac{a_0 b_i}{\av{b}N^2}
      \right)
    \left( 1 - \frac{a_0 b_j}{\av{b}N^2}
       \right)
  \right]^{T N }\nonumber \\
&\simeq & \exp\left[ - \lT'(b_i + b_j)\right],
\nonumber
\end{eqnarray}
where we have defined the new parameter $\lT' = a_0T/\av{b}N$ and, in
the last step of the previous expressions, we have performed and
expansion for large $N$ a finite $T$. From here, we obtain
$\Pi'_T(i,j) = 1- Q'_T(i,j)$. As we can see, this modified model is
exactly mappable to the activity driven model (see Eq.~(\ref{eq:7}),
with the simple translation $\lT \to \lT'$; all the general expression
derived above hold thus in this case, and can be worked out, upon
providing the appropriate expression for the \textit{attractiveness}
distribution $F(b)$.

\section{Conclusions}
\label{sec:conclusions}

The activity driven model represents an interesting approximation to
temporal networks, providing an preliminary explanation of the origin
of the degree distribution of integrated social networks, in terms of
the heterogeneity of the agents' activity, and the distribution of
this quantity. Here we have explored the full relation between
topology and activity distribution, obtaining analytical expressions
for several topological properties of the integrated social networks
for a general activity potential, in the thermodynamic limit of large
number of agents, $N\rightarrow \infty$, and finite integration time
$T$.  To tackle this issue, we have applied the hidden variables
formalism, by mapping the aggregated network to a model in which the
probability of connecting two nodes depends on the hidden variable (in
this case represented by the activity potential) of those nodes.  Our
analysis is complemented by numerical simulations in order to check
theoretical predictions against concrete examples of activity
potential distributions.  Using our formalism, we can demonstrate
rigorously that the integrated degree distribution at time $T$ takes
the same functional form as the activity potential distribution, as a
function of the rescaled degree $k/T-\av{a}$. This is however an
asymptotic result, which is well fulfilled for an activity potential
power-law distributed, as empirically measured in a wide range of
social interaction settings, which fails for simple constant or
homogeneous distributions. We also show that the aggregated networks
show in general disassortative degree correlations, at odds with the
assortative mixing revealed in real social networks.  The clustering
coefficient is low, $\av{c} \sim T/N$, comparable with a random
network.  

Our study opens interesting direction for future work, concerning, for
example, the clarification of role of integration time in the
properties of dynamical process on activity driven networks, and the
possible modifications of the activity driven network model, in order
to incorporate some properties of real social networks currently
missed, such as a high clustering coefficient, assortative mixing by
degree or a community structure \cite{Newman2010}.

\begin{acknowledgments}
  We acknowledge financial support from the Spanish MICINN, under
  project No. FIS2010-21781-C02-01. R.P.-S. acknowledges additional
  financial support from ICREA Academia, funded by the Generalitat de
  Catalunya
\end{acknowledgments}


\bibliographystyle{apsrev4-1}


\bibliography{Bibliography.bib}

\end{document}